# Authorship analysis of specialized vs diversified research output[1]

Giovanni Abramo

*Laboratory for Studies in Research Evaluation*
*at the Institute for System Analysis and Computer Science (IASI-CNR)*
*National Research Council of Italy*
ADDRESS: Istituto di Analisi dei Sistemi e Informatica, Consiglio Nazionale delle Ricerche, Via dei Taurini 19, 00185 Roma - ITALY
giovanni.abramo@uniroma2.it

Ciriaco Andrea D'Angelo

*University of Rome "Tor Vergata" - Italy and*
*Laboratory for Studies in Research Evaluation (IASI-CNR)*
ADDRESS: Dipartimento di Ingegneria dell'Impresa, Università degli Studi di Roma "Tor Vergata", Via del Politecnico 1, 00133 Roma - ITALY
dangelo@dii.uniroma2.it

Flavia Di Costa

*Research Value s.r.l.*
ADDRESS: Research Value, Via Michelangelo Tilli 39, 00156 Roma- ITALY
flavia.dicosta@gmail.com

**Abstract**
The present work investigates the relations between amplitude and type of collaboration (intramural, extramural domestic or international) and output of specialized versus diversified research. By specialized or diversified research, we mean within or beyond the author's dominant research topic. The field of observation is the scientific production over five years from about 23,500 academics. The analyses are conducted at the aggregate and disciplinary level. The results lead to the conclusion that in general, the output of diversified research is no more frequently the fruit of collaboration than is specialized research. At the level of the particular collaboration types, international collaborations weakly underlie the specialized kind of research output; on the contrary, extramural domestic and intramural collaborations are weakly associated with diversified research. While the weakness of association remains, exceptions are observed at the level of the individual disciplines.

# 1. Introduction

The ever more collaborative nature of scientific research is seen in the ongoing increase in the share of multi-authored papers, observed in the majority of scientific fields (Wuchty, Jones, & Uzzi, 2007; Leahey, 2016). As stated by Siciliano, Welch, and Feeney (2017), "the production of scientific knowledge is an inherently social process, making professional networks important for producing science outcomes". The choices and strategies involved in collaboration have been thoroughly studied in the literature (Bozeman & Corley, 2004; Katz & Martin, 1997). Bammer (2008) provides a list of motivations: "access to expertise or particular skills, access to equipment or resources, cross-fertilization across disciplines, improved access to funding, learning tacit knowledge about a technique, obtaining prestige, visibility or recognition, and enhancing student education".

The increasing complexity of scientific challenges has led to greater specialization, and at the same time to a need to bring together different competencies: a necessity that becomes ever more critical as scientists continue to specialize. To address the increasing complexity of their own disciplinary area, scientists tend towards collaboration with colleagues in the same field (specialized research). At the same time, the multi-disciplinary character of many problems induces collaboration among scholars of different areas (Boh, Ren, Kiesler, & Bussjaeger, 2007; Wray, 2005; Clark & Llorens, 2012; Jones, Wuchty, & Uzzi, 2008). For one or more of the members of the multi-disciplinary teams that are formed, this often leads to diversification beyond the boundaries of their own disciplines. We could think of examples such as collaborations between doctors and statisticians in epidemiological studies, or between mathematicians and economists in financial modeling. However, diversification in research also arises from other causes than involvement in multidisciplinary research projects. A well known example in the field of bibliometrics is that of the physicist Jorge Hirsch: with a substantial history of publication in physics, he then independently wrote a pair of publications introducing the now-famous "h-index" (Hirsch, 2005, 2007).

The study of the relation between scientific specialization/diversification of individual scientists and the different types of collaborations can be of interest to both policy makers and the scientists themselves. In fact in the daily practice of scientific activity, "The outcome of an interaction is to a large extent dependent on the balancing of contradicting motivations accompanying the establishment of collaborations" (Mayrose & Freilich, 2015). The intensity with which these collaborations develop (intramural/extramural, domestic/international, intradisciplinary/interdisciplinary) varies on the basis of contextual and personal factors, such as gender (Bozeman & Gaughan, 2011; Abramo, D'Angelo, & Murgia, 2013a; Ozel, Kretschmer, & Kretschmer, 2014), age (van Rijnsoever & Hessels, 2011; Aschhoff & Grimpe, 2011; Abramo, D'Angelo, & Murgia, 2017), academic rank (Lee & Bozeman, 2005; Rivellini, Rizzi, & Zaccarin, 2006; Abramo, D'Angelo, & Murgia, 2014), the research discipline involved (Yoshikane & Kageura, 2004; Gazni, Sugimoto, & Didegah, 2012). Even within a single discipline there can be notable heterogeneity in the forms of activating collaboration, due to the different specializations involved (Piette & Ross, 1992; Newman, 2001; Moody, 2004; Abramo, D'Angelo & Murgia, 2013b). Studies by Cole and Zuckerman (1975), and more recently by Moody (2004), have demonstrated the existence of a positive relation between the growth of specialty areas and collaboration rates.

Various studies have investigated the effect of interdisciplinary collaborations on the impact of the research products (Steele and Stier, 2000; Rinia et al., 2001; Levitt and Thelwall, 2008; Larivière and Gingras, 2010; Yegros-Yegros, Rafols, & D'Este, 2015; Wang, Thijs, & Glänzel, 2015; Chen et al., 2015; Abramo, D'Angelo, & Di Costa, 2017d). However the results are often contrasting, in part because of resorting to different indicators for measuring interdisciplinary research (Wang, Thijs, & Glänzel, 2015; Abramo, D'Angelo, & Di Costa, 2017d).

Bibliometric studies on research diversification have appeared only very recently. A few scholars have provided methodological contributions on identifying the field of specialization of a researcher (Abramo, D'Angelo & Di Costa, 2017a; Mizukami, Mizutani, Honda, Suzuki & Nakano, 2017). As well, Abramo, D'Angelo & Di Costa, (2017b) tested the hypothesis that to obtain high impact results, a scientist must develop a certain level of specialization and remain focused on their own dominant research topic. Their analyses showed that it is clearly more difficult to obtain important results in knowledge domains outside the area of one's core competence. Having established the link between specialization/diversification and scientific impact, Abramo and D'Angelo (2017a) then verified the existence of gender differences in research diversification, and subsequently the effects of age and academic rank on propensity to diversify (Abramo, D'Angelo & Di Costa, 2017c). Finally, Abramo, D'Angelo, and Di Costa (2018) found that a scientist's output resulting from research diversification is more often than not the result of collaborations with multidisciplinary teams.

In the current of these works, we now propose to study the relations between the character of a publication (specialized or diversified – meaning within or beyond the author's dominant research topic) and some characteristics of the research team that produced it, in particular the numerosity of the team and the type of collaboration involved (international, domestic, extramural, intramural). In particular, we intend to respond to the following research questions:

- *Are specialized publications more frequently the fruit of collaboration than diversified ones?*
- *Among intramural, extramural and international types of collaboration, which are more associated with the specialized or diversified forms of publication by an author?*
- *On average, do specialized publications show a significantly higher number of co-authors than diversified ones?*
- *Are there differences across disciplines?*

To respond to these questions we will conduct statistical analyses on a dataset comprising the 2012-2016 scientific production indexed in the Web of Science, as authored by all Italian professors in the sciences (nearly 23,500 scientists).

Results are relevant at both government and management level. With aims of increasing the pace of scientific and technological advancements, boosting research productivity and addressing complex societal problems, a growing number of national governments and research organizations are making efforts to foster research collaboration, especially international, and interdisciplinary research, which often underlines research diversification. Shedding light on the relationship between the different types of research collaboration and diversification/specialization may help formulate coherent policies and synergistic initiatives to actualize them. Furthermore, there is a link between: i) research collaboration and performance of the individual scientist, although the causal nexus between the two has still not been fully clarified (Lee & Bozeman, 2005; He, Geng, & Campbell-Hunt, 2009; Ynalvez & Shrum, 2011; Abramo, D'Angelo, & Murgia, 2017); and ii) research diversification and performance (Abramo, D'Angelo & Di Costa, 2017b). Finally, because, all others being equal, the higher the number of authors per publication, the lower the research productivity of individuals (Abramo & D'Angelo, 2014), whenever interdisciplinary research implies diversification, if diversified publications show a significantly higher number of co-authors than specialized one, policy makers may expect resistance to policy incentives aimed at interdisciplinary research.

The next section presents the dataset and describes the methodological details of the analyses. Sections 3 provides the results, Section 4 offers the conclusions and authors' considerations.

## 2. Data and methods

### 2.1 Data

We observe the 2012-2016 scientific production,[2] indexed in WoS and authored by Italian university professors in the sciences, for purposes of verifying the existence of a relation between the character of a publication (specialized or diversified) and the type of collaboration that produced it. We have chosen to exclude social sciences, because it turns out that Italian professors in these fields have a too large share of their publication output in sources that are not covered by WoS.[3] Findings then would not be robust enough. It must be noted that the faculty staff in the sciences represents above 60% of total staff, while social scientists only 20%. In arts & humanities the situation is even worse.

In the Italian university system all academics are classified in one and only one field, named scientific disciplinary sector (SDS), of which there are 370 in all. SDSs are grouped into disciplines, named university disciplinary areas (UDAs), of which there are 14. The sciences are composed of 9 UDAs (Mathematics and computer sciences, Physics, Chemistry, Earth sciences, Biology, Medicine, Agricultural and veterinary sciences, Civil engineering, Industrial and information engineering); within these UDAs there are 192 SDSs. The Italian Ministry of Education, University and Research, maintains a database which indexes all Italian academics and provides information on their university affiliations, field classification, academic rank, and gender.[4] Given a publication indexed in WoS (article, review article and conference proceeding), we are able to apply an algorithm to disambiguate the true identity of the authors, so that each

---

[2] A five-year publication period is adequate to reduce the problem of paucity of publications at field level and year-dependent fluctuations with systematic effects on results (Abramo, D'Angelo, & Cicero, 2012).
[3] The percentages of Italian social science professors (by field) who have none of their 2001-2004 outputs covered by WoS, are: Political economy, 66.2%; Economic policy, 75.0%; Finance, 69.2%; History of economic thought, 86.7%; Econometrics, 28.0%; Applied economics, 77.4%; Business administration, 96.0%; Corporate finance, 87.2%; Financial management, 100.0%; Business organisation, 81.4%; Economics of financial intermediaries, 95.3%; Economic history, 95.3%; Commodity studies, 67.9% (D'Angelo & Abramo, 2015).
[4] http://cercauniversita.cineca.it/php5/docenti/cerca.php, last access 20 February 2019.

publication is attributed to the academics that produced it (D'Angelo, Giuffrida, & Abramo, 2011).[5]

The Italian professors in the 192 Science SDSs included in the analysis are those whose 2012-2016 publication portfolio meet (for reasons of significance) the following requirements:

- at least 5 publications are WoS indexed;
- the publication portfolio falls in at least 2 different WoS subject categories (SCs);
- the publication portfolio presents one prevalent SC.[6]

As will be seen in Section 2.3, the prevalent SC is the one with the largest number of the professor's publications. Furthermore, the attribute of "specialized" or "diversified" is not a characteristic of the publications themselves. Instead, we identify the character of the work considering the prevalent SC of the author that produced it: for example the same publication, having two co-authors, could be "specialized" for the first and "diversified" for the second.

The dataset thus prepared consists of 23,486 professors, who produced a total of 255,800 WoS-indexed publications. They are distributed among UDAs as shown in Table 1.

*Table 1: Dataset of the analysis, by UDA*

| UDA* | SDSs | Professors | Publications | Authorship |
|---|---|---|---|---|
| 1 | 10 | 1662 | 17317 | 23267 |
| 2 | 8 | 1771 | 25084 | 101595 |
| 3 | 12 | 2361 | 28173 | 54336 |
| 4 | 12 | 698 | 7088 | 10897 |
| 5 | 19 | 3307 | 35844 | 58844 |
| 6 | 50 | 6686 | 85367 | 175833 |
| 7 | 30 | 2006 | 17511 | 32967 |
| 8 | 9 | 964 | 10511 | 15818 |
| 9 | 42 | 4031 | 60137 | 98154 |
| Total | 192 | 23486 | 255800† | 571711 |

\* 1, Mathematics and computer science; 2, Physics; 3, Chemistry; 4, Earth sciences; 5, Biology; 6, Medicine; 7, Agricultural and veterinary sciences; 8, Civil engineering; 9, Industrial and information engineering

## 2.2 Defining the collaboration patterns through co-authorship analysis

The research production function varies across disciplines. Production factors, i.e. the combination of labor and capital, vary as well. Collaboration natures and norms tend to be different too. For instance, humanists highly value sole-authorship, while physicians are more likely to be involved in collaborative projects. Abramo and D'Angelo (2017b) showed that in the sciences, the average number of co-authors per paper is lowest in mathematics (2.74) and highest in medical sciences (6.13).[7] To help readers have a better understanding of the collaboration behavior across disciplines, and in Italy in particular, we refer them to the work by Abramo, D'Angelo and Murgia (2013b). The authors provide an exhaustive analysis of variation in research collaboration patterns across disciplines and in fields within disciplines.

---
[5] The harmonic average of precision and recall (F-measure) of authorships, as disambiguated by the algorithm, is around 97% (2% margin of error, 98% confidence interval).

[6] For the purpose of the current work, a scholar "specialized" in more than one SC is not specialized by convention.

[7] Mathematics and computer science, 2.74; Physics, 4.54; Chemistry, 4.83; Earth sciences, 4.03; Biology, 5.07; Medicine, 6.13; Agricultural and veterinary sciences, 4.52; Civil engineering, 2.87; Industrial and information engineering, 3.47.

Operationally, to identify the types of collaboration from the publication byline, we use the taxonomy of Abramo, D'Angelo, and Murgia (2013b). In particular, a publication is classified as resulting from collaboration (*cp*) if the publication byline shows two or more co-authors. Then for each professor of the publication in co-authorship, we can identify the possible presence of three types of collaboration, based on whether at least one other co-author belongs to:
- a foreign organization (international collaboration – *cefp*);
- a domestic organization different from his or her own (extramural domestic collaboration – *cedp*); and
- the same university (intramural collaboration – *cip*).

Thus, the set of all collaboration (*cp*) is a superset of three forms of collaboration (*cip*, *cedp*, *cefp*).

Table 2 shows the examples of two publications and their relative bylines. The first shows six authors. Only three of these are assigned to our dataset (see Section 2.1): Rigano Daniela of the University of Naples 'Federico II', and Bruno Maurizio and Rosselli Sergio, both of the University of Palermo.

In this first byline, for Daniela Rigano we identify the following types of collaboration:
- international (cefp=1), for the presence of a Turkish university in the address list;
- extramural domestic (cedp=1), for the presence of the University of Palermo in the address list;
- intramural (cip=1), since there are co-authors affiliated with the same University (Naples 'Federico II'), although we can observe these are non-faculty staff.

Given similar observations for co-authors Bruno Maurizio and Rosselli Sergio, in this case the publication in question is again considered the fruit of three types of collaboration.

The second example is a publication co-authored by three scientists, all disambiguated Italian professors and consequently included in the dataset. In this case, for the first two (both affiliated with Turin Polytechnic) the publication is identified as the fruit of both intramural and extramural domestic collaborations. For the third author (affiliated with the University of Turin) it is only the result of an extramural domestic collaboration.

Of course these operational definitions of collaborations totally ignore the role of the author in the collaborating team. This is an evident limit of any large-scale bibliometric analysis.

Table 3 presents descriptive statistics on collaboration patterns, in output per UDA. We observe that 99% of the authorships in the dataset (last row of Table 3) are the fruit of collaboration, and in particular:
- 42.2% arise from international collaborations (with a maximum in Physics, 84.4%, and a minimum in Medicine, 29.5%);
- 61.8% arise from extramural domestic collaborations (with a maximum in Physics, 90.7%, and a minimum in Industrial and information engineering, 35%);

- 86.2% arise from intramural collaborations (with a maximum in Physics, 93.2%, and a minimum in Mathematics and computer science at the bottom, 61.2%).

*Table 2: Examples of different types of collaboration*

| WoS_id | Author list | Address list |
|---|---|---|
| 257581600002 | Formisano C; Rigano D; Senatore F; Celik S; Bruno M; Rosselli S | Univ Naples Federico II, Dipartimento Chim Sostanze Nat, NAPLES, I-80131, ITALY; Canakkale Onsekiz Mart Univ, Fac Sci & Literature, Dept Biol, CANAKKALE, TURKEY; Univ Palermo, Dipartimento Chim Organ, PALERMO, I-90128, ITALY |
| 186442700012 | Grosso A; Dellacroce F; Tadei R | Univ Turin, Dipartimento Informat, TURIN, I-10149, ITALY; Polytech Univ Turin, DIPARTIMENTO AUTOMAT & INFORMAT, TURIN, I-10129, ITALY |

*Table 3: Authorship statistics of the publications in the dataset, by UDA*

| UDA* | Authorship | In collaboration | International | Extramural domestic | Intramural |
|---|---|---|---|---|---|
| 1 | 23267 | 94.7% | 42.8% | 42.3% | 61.2% |
| 2 | 101595 | 99.1% | 84.4% | 90.7% | 93.2% |
| 3 | 54336 | 99.4% | 38.7% | 56.3% | 89.4% |
| 4 | 10897 | 98.8% | 47.7% | 66.0% | 74.6% |
| 5 | 58844 | 99.4% | 39.4% | 64.6% | 87.2% |
| 6 | 175833 | 99.6% | 29.5% | 67.1% | 84.3% |
| 7 | 32967 | 99.5% | 31.2% | 52.9% | 88.8% |
| 8 | 15818 | 97.1% | 27.7% | 35.2% | 80.9% |
| 9 | 98154 | 98.5% | 30.0% | 35.0% | 87.3% |
| Total | 571711 | 99.0% | 42.2% | 61.8% | 86.2% |

* 1, Mathematics and computer science; 2, Physics; 3, Chemistry; 4, Earth sciences; 5, Biology; 6, Medicine; 7, Agricultural and veterinary sciences; 8, Civil engineering; 9, Industrial and information engineering

**2.3 Specialized vs diversified research output**

The classification of a scientist's publications requires identification of their prevalent field of research, for which we use the procedure of Abramo, D'Angelo and Di Costa (2017b).

The procedure is as follows: i) we identify the scientific production of the scientist over a period of time; ii) we associate the publications indexed in the Web of Science (WoS) with the subject categories (SCs) of the hosting journal; iii) we identify the SC (prevalent field) with the largest number of the scientist's publications. Such SC represents the "prevalent" field of research of the scientist, and the publications falling in it are termed as "specialized" publications. Table 4 shows the case of John Doe, professor in experimental physics. Over the five years considered, his scientific production amounted to 32 publications. The analysis of the SCs associated with the journals hosting these works reveals that the most frequent is "Optics", which repeats for 24 works, either alone or with other SCs. We conclude that scientific production of this author consists of 32 publications, of which 24 specialized and 8 diversified. We then refer to the products in a scientist's prevalent field as "specialized publications", and to the remainder as "diversified publications".[8]

Table 5 reviews the entire dataset, showing that 61.6% of total authorships relate to specialized publications and 38.4% to diversified publications. Of these shares, 0.8% of the specialized publications are single-authored papers, compared to 1.1% for the diversified ones. Physics is the only UDA where diversified single-authored papers are more frequent than specialized ones.

*Table 4: Scientific production of an Italian experimental physics professor*

| Journal | No. of publications | Subject categories of the journal | Specialized |
|---|---|---|---|
| Journal of Biomedical Optics | 8 | Biochemical Research Methods; Optics; Radiology, Nuclear Medicine & Medical Imaging | Yes |
| Applied Optics | 6 | Optics | Yes |
| Optics Express | 6 | Optics | Yes |
| Physics In Medicine and Biology | 5 | Engineering, Biomedical; Radiology, Nuclear Medicine & Medical Imaging | No |
| Optics Letters | 4 | Optics | Yes |
| Postharvest Biology and Technology | 3 | Agronomy; Food Science & Technology; Horticulture | No |

*Table 5: Share of specialized (diversified) authorships in the dataset, by UDA (percentage of single authored papers in parentheses)*

| UDA* | Authorship | Specialized | Diversified |
|---|---|---|---|
| 1 | 23267 | 68.9% (6.2%) | 31.1% (3.5%) |
| 2 | 101595 | 72.0% (0.7%) | 28.0% (1.3%) |
| 3 | 54336 | 53.4% (0.7%) | 46.6% (0.5%) |
| 4 | 10897 | 59.3% (1.4%) | 40.7% (0.9%) |
| 5 | 58844 | 50.2% (0.8%) | 49.8% (0.5%) |
| 6 | 175833 | 58.4% (0.5%) | 41.6% (0.3%) |
| 7 | 32967 | 60.7% (0.6%) | 39.3% (0.5%) |
| 8 | 15818 | 63.5% (3.1%) | 36.5% (2.5%) |
| 9 | 98154 | 66.7% (1.6%) | 33.3% (1.3%) |
| Total | 571711 | 61.6% (1.1%) | 38.4% (0.8%) |

*\* 1, Mathematics and computer science; 2, Physics; 3, Chemistry; 4, Earth sciences; 5, Biology; 6, Medicine; 7, Agricultural and veterinary sciences; 8, Civil engineering; 9, Industrial and information engineering*

## 2.4 Statistical methods

In order to respond to the research questions, we use two types of statistical analyses
    a) One-way hypothesis test;
    b) Logistic regression.

The first type of analysis allows to assess whether there is an association between the specialized/diversified character of a publication of a given author and each of the four collaboration types attested by its byline. Given the dichotomous trait of the variables at play, we use: a) the Pearson's $\chi^2$ test and, b) an effect size measure, the Cramér's V which, in our case, coincides with the phi coefficient, since we deal with 2x2 matrices (Cramér, 1946). The Pearson's $\chi^2$ gives us an indication of the statistical significance of the association, while the Cramer's V, controlling for the sample size, measure the strength of the association and its practical significance. As for the number of co-authors (a continuous variable), we use the Mann-Whitney two-sample statistic form of the Wilcoxon rank-sum test, still valid for data from any distribution, whether normal or not, and rather insensitive to outliers (Mann & Whitney, 1947).

---

[8] An alternative way to judge whether a scientist's output is specialized or not is by looking at the content of individual papers by text-mining techniques.

Finally, we apply the logistic regression to test the simultaneous effect of all the five independent variables on the trait (specialized/diversified) of a publication. In this regard, we opt for a binomial logistic regression, given the binary feature of the dependent variable Y ("0" for diversified publications; "1" for specialized ones).

## 3. Analysis and results

In this section we provide answers to the research questions. We start by analysing the data at aggregate overall level. Then, we repeat the analysis at the discipline level, to identify possible differences across disciplines.

### 3.1 Aggregate level analysis

In Table 5, we showed that 61.6% of total authorships relate to specialized publications and 38.4% to diversified publications. Table 6 illustrates the contingency matrix with the breakdown of such subsets by the *cefp* variable. Among the specialized publications, 45.1% (158969/352423) are the result of collaboration with foreign organization; among diversified publications, they are 37.5% (82237/219288). Comparing observed data with "expected" ones, one obtains a value of Pearson $\chi^2$ (3200.0), and a p value (0.000) confirming the statistical significance of the association between the two variables. However, the Cramér's V value (0.0749) indicates a very weak association (Rea & Parker, 1992), moreover positive. That means that at the aggregate level specialized publications result from international collaborations more likely than diversified publications, although the association between the two variables is weak.

*Table 6: Contingency matrix for publications in the dataset. Influence of international collaborations*

|   |   | *cefp*\* | | |
|---|---|---|---|---|
|   |   | 0 | 1 | Total |
| Y† | 0 | 137051 | 82237 | 219288 |
|   | 1 | 193454 | 158969 | 352423 |
|   | Total | 330505 | 241206 | 571711 |

† 1, for specialized publications; 0, for diversified ones
\* 1, for publications resulting from collaboration with foreign organizations; 0, otherwise.
Pearson $\chi^2$ = 3200; p value = 0.000; Cramér's V = 0.0749

We repeat the same analysis for the different types of collaboration. Results (Table 7) show a significant, very weak and negative relation between each type of collaboration and the specialization of resulting publications with the exception of *cefp* showing a positive association.

*Table 7: One-way test of hypotheses concerning the association of collaboration type and the trait specialized vs diversified of resulting publications*

|  | Pearson $\chi^2$ | Cramér's V | Sign |  |
|---|---|---|---|---|
| cefp | 3200.0 | 0.075 | (+) | *** |
| cedp | 18.1 | 0.006 | (-) | *** |
| cip | 130.9 | 0.015 | (-) | *** |
| cp | 136.0 | 0.015 | (-) | *** |
| No. of authors | n.a. | n.a. | (-) | *** |

*\* p < 0.1; \*\* p < 0.05; \*\*\* p < 0.01*
*cp, publication in collaboration; cip, publication in intramural collaboration; cedp, publication in extramural domestic collaboration; cefp, publication in international collaboration.*

The logistic regression, simultaneously considering the effect of all independent variables, could provide a more clear answer to the research questions.

We have conducted the regression analysis for the aggregate dataset, but after running a postregression test for VIF, we found that the value for *cp* is above 10 (11.9). We have therefore decided to drop *cp*. The regression analysis, without this variable,[9] is shown in Table 8 with both the logit coefficients and odds ratios (OR): the latter represent the odds that Y=1 (meaning the publication would be specialized) when variable X increases by one unit (meaning, for example, that the publication would be the result of collaboration compared to the case that it is not). The value of the Prob ($\chi^2$) test, at less than 0.05, indicates that the model has relevant explanatory power.

*Table 8: Logistic regression analysis*

| Variable | Logit coefficients | Odds Ratio | p-values |
|---|---|---|---|
| cefp | 0.126 | 1.134 | 0.000*** |
| cedp | -0.176 | 0.839 | 0.000*** |
| cip | -0.174 | 0.840 | 0.000*** |
| No. of authors | 0.000 | 1.000 | 0.000*** |
| const | 0.597 |  | 0.000*** |

*Y=1 for specialized publication; Y=0 for diversified publication*
*\* p < 0.1; \*\* p < 0.05; \*\*\* p < 0.01*
*Prob ($\chi^2$) = 0.000. N=569350 (co-authorship)*
*cip, publication in intramural collaboration; cedp, publication in extramural domestic collaboration; cefp, publication in international collaboration.*

We observe that all variables are statistically significant. Therefore, we can conclude that the specialized or diversified trait of a publication of a given author depends on the type of collaboration underlying the publication.

The only OR greater than 1 is seen for the variable *cefp*. We deduce that it is more likely (13.4% of probability) for a specialized publication to arise from collaboration with foreign institutions. On the contrary, those arising from intramural or extramural national collaborations are less likely to be specialized (-16.0 and -16.1% respectively). The number of authors does not seem to significantly affect the trait of the publication (logit coefficients are practically nil).

In conclusion, the specialized character of a publication by an author is more probable in the works that are the fruit of collaboration with foreign partners and less probable in those that arise from domestic collaborations (both intramural and extramural). Finally, the number of authors impacts very marginally on the probability of a publication being specialized.

---

[9] The post regression test for VIF now shows values always below 3 for all variables.

It is to be noted (Table 5) that the dataset is dominated by the UDA medicine, which accounts for about 31% of all observations (co-authorship). Thus, the overall outcome is evidently dependent on the structure of the data sample, which makes it necessary to repeat the analysis at a lower level of aggregation, that of the individual disciplines.

**3.2 Discipline level analysis**

As for the first research question, verifying whether specialized publications are more (or less) frequently the fruit of collaboration than diversified ones, the second column of Table 9 shows the results of the one-way tests of hypothesis.

At UDA level the relation results as significant in all disciplines, but Agricultural and veterinary sciences (UDA7). The strength of the relation is very weak (Cramer's V less than 0.1) and always negative, but in Physics (UDA 2). We can conclude that diversified publications are more frequently the fruit of collaboration, although the absolute value of the differences is quite low.

For the second research question, concerning which types of collaboration are more frequently associated with specialized or diversified types of publication, results are shown in column 3-5 of Table 9. Substantial differences emerge across disciplines. In regards to the *cefp* variable: however weakly, foreign collaboration is positively associated with specialization in all UDAs, but UDAs 7, 8, and 9 whose values are not significant.

The results from this analysis are similarly differentiated concerning the other two types of collaboration (extramural domestic and intramural). Keeping in mind the very weak association between variables, as shown by the Cramer's V, in some UDAs the diversified character seems significantly more frequent among publications that are the fruit of both these forms of collaboration (in Mathematics, Earth sciences, Medicine, Industrial and information engineering), while the opposite is true in Physics. In Agricultural and veterinary sciences, the association of the collaboration type to the character of the resulting publications is not univocal: positive per intramural and negative for extramural domestic. For Chemistry and Biology the only significant association concerns the diversified nature of publications arising from extramural domestic collaborations.

Finally, concerning the third research question, verifying whether specialized publications on average have a higher number of co-authors than diversified ones, the results of the Mann-Whitney two-sample test are presented in the last column of Table 9. They indicate a consistently significant relation (p-values always less than 0.05), but positive only for UDA 2, Physics.[10] In all UDAs but this one exception, we can conclude that the number of authors is significantly less for specialized publications than for diversified ones.

---

[10] High energy physics and astrophysics research is characterized by multinational collaborations involving very large research teams. This is a clear example of a focused, collaborative, international structure that is rare in other areas.

*Table 9: One-way test of hypotheses concerning specialized vs diversified publications. Cramer's V statistics for binary variables (cp, cefp, cedp, cip); Mann-Whitney p values, for "No. of authors". Sign of relation in parentheses.*

| UDA* | cp | cefp | cedp | cip | No. of authors |
|---|---|---|---|---|---|
| 1 | 0.054(-)*** | 0.028(+)*** | 0.045(-)*** | 0.099(-)*** | 0.000(-)*** |
| 2 | 0.027(+)*** | 0.194(+)*** | 0.119(+)*** | 0.046(+)*** | 0.000(+)*** |
| 3 | 0.014(-)*** | 0.013(+)*** | 0.032(-)*** | 0.002(+) | 0.000(-)*** |
| 4 | 0.021(-)** | 0.041(+)*** | 0.021(-)** | 0.038(-)*** | 0.000(-)*** |
| 5 | 0.014(-)*** | 0.031(+)*** | 0.026(-)*** | 0.006(-) | 0.000(-)*** |
| 6 | 0.014(-)*** | 0.030(+)*** | 0.009(-)*** | 0.05(-)*** | 0.000(-)*** |
| 7 | 0.007(-) | 0.002(-) | 0.057(-)*** | 0.025(+)*** | 0.000(-)*** |
| 8 | 0.019(-)** | 0.009(+) | 0.007(-) | 0.035(-)*** | 0.000(-)*** |
| 9 | 0.011(-)*** | 0.004(+) | 0.068(-)*** | 0.006(-)* | 0.000(-)*** |

*1 - Mathematics and computer science; 2 - Physics; 3 - Chemistry; 4 - Earth sciences; 5 - Biology; 6 - Medicine; 7 - Agricultural and veterinary sciences; 8 - Civil engineering; 9 - Industrial and information engineering.*
*cp - publication in collaboration; cip - publication in intramural collaboration; cedp - publication in extramural domestic collaboration; cefp - publication in international collaboration.*
*\* p < 0.1; \*\* p < 0.05; \*\*\* p < 0.01*

As for the logistic regression, we excluded the variable *cp* also at UDA level, because of multicollinearity, as revealed by the post-regression tests run for each UDA. The outcomes of the logistic regression in Table 10 prove that concerning the international co-authorship variable (*cefp*), the OR values are statistically significant in all UDAs but Mathematics and computer science, and greater than 1 in five cases (Physics; Chemistry; Earth sciences; Biology; and Medicine). Physics (UDA 2) registers the highest increase (94.1%) in the probability that specialized publications are the fruit of international collaborations.

Instead, as for the extramural domestic variable (*cedp*), the OR values are statistically significant in all UDAs but always less than 1 except in UDA 2 (physics). Thus, with the sole exception of Physics, it seems that the specialized kind of publication occurs with less probability in works that are the fruit of extramural domestic collaboration. The extreme case is represented by publications in industrial and information engineering: the probability that specialized publications are the fruit of extramural domestic collaboration is 25.8% lower than diversified.

For the intramural variable (*cip*), the OR values are statistically significant in seven out of nine UDAs and greater than 1 in Agricultural and veterinary sciences where the specialized kind of publication is more characteristic (+11.8% in probability) of collaboration with colleagues from the same institution.

Finally, concerning the number of authors, the values of OR are statistically significant in all UDAs

In order to verify the robustness of this analysis we carried out other regressions varying thresholds for the dataset selection. In particular, we varied the minimum share of publications in the prevalent SC and the distance between the first and the second SC ranked for number of publications. We repeated the regressions of Table 10 for four different scenarios: details are shown in the appendix and reveal a truly satisfactory alignment of outcomes both in terms of significance of ORs and of their sign, in all UDAs.

*Table 10: Logistic regression analysis (odds ratios), by UDA*

| UDA† | Obs | cefp | | cedp | | cip | | No. of authors | | Prob (χ2) |
|---|---|---|---|---|---|---|---|---|---|---|
| 1 | 22,990 | 0.959 | ns | 0.726 | *** | 0.584 | *** | 1.001 | *** | 0.000 |
| 2 | 101,333 | 1.941 | *** | 1.265 | *** | 0.923 | *** | 1.000 | *** | 0.000 |
| 3 | 54,247 | 1.038 | *** | 0.877 | *** | 1.013 | ns | 1.001 | *** | 0.000 |
| 4 | 10,860 | 1.159 | * | 0.931 | * | 0.863 | *** | 0.997 | * | 0.000 |
| 5 | 58,661 | 1.108 | *** | 0.895 | *** | 0.977 | ns | 1.001 | *** | 0.000 |
| 6 | 175,266 | 1.092 | *** | 0.933 | ** | 0.766 | *** | 1.000 | ** | 0.000 |
| 7 | 32,877 | 0.988 | *** | 0.794 | * | 1.118 | *** | 1.001 | * | 0.000 |
| 8 | 15,756 | 0.986 | ** | 0.923 | *** | 0.805 | *** | 1.002 | *** | 0.000 |
| 9 | 97,360 | 0.999 | *** | 0.742 | *** | 0.911 | *** | 1.000 | *** | 0.000 |

† *1 - Mathematics and computer science; 2 - Physics; 3 - Chemistry; 4 - Earth sciences; 5 - Biology; 6 - Medicine; 7 - Agricultural and veterinary sciences; 8 - Civil engineering; 9 - Industrial and information engineering.*
*Y=1 for specialized publication; Y=0 for diversified publication*
*\* p < 0.1; \*\* p < 0.05; \*\*\* p < 0.01*
*cip - in intramural collaboration; cedp - in extramural domestic collaboration; cefp - in international collaboration.*

## 4. Conclusions

The present work continues a stream of inquiry, initiated by the authors, concerning diversification versus specialization in research activity at the individual level. In this case the work concentrates on the amplitude and type of collaboration, where existent, that underlies the specialized or diversified kinds of research output. The study of the connection between collaboration and diversification/specialization can be important for policies and initiatives aimed at fostering one or the other, by informing policy makers on the form of collaboration more associated with each one.

The objective of the present work was to determine if and how "specialized" publications differentiate from "diversified" ones in function of the characteristics of the team that generates them, in particular the number of authors and the type of collaboration developed (intramural, extramural domestic or international). Given the dichotomous character of the "response" variable, we have used a logit model.

We observe that the association between the type of collaboration and the spezialized/diversified character of research output, is very weak at both overall and disciplinary levels. Although weakly, specialized output as compared to the diversified one, is more associated with collaboration involving foreign institutions, and less with extramural and intramural domestic collaborations or with numerosity of co-authors (Table 8). These results are in line with what we would expect. The periodic international congresses of the epistemic communities favor occasions for meeting and exchange of ideas between scholars of different nations within the same field of research, and so the birth of "international" research projects of shared interest. Instead there is less probability of opportunities for meeting and exchange at international level between scholars of different fields, which could induce some of the scholars towards field diversification for participation in research projects, evidently of multi-disciplinary character, with colleagues from other fields. It seems more likely that such dynamics would occur at domestic level in general, and through intramural encounters in particular. Here, geographic proximity can above all favor the possibility of understanding who has competences different from one's own and necessary for a research project under consideration. Not only, it can also favor the encounters and exchanges necessary for a scholar to be convinced that it is worth the risk of applying their knowledge in different fields.

The few exceptions noted at the discipline level (Table 10), for each type of collaboration, give reason for reflection, since they necessarily call into question the specific natures and the distinctive behavior of the scientists of these disciplines. We observe though similar behavior in Chemistry, Earth sciences, Biology and Medicine. Keeping always in mind that the association is weak, in these UDAs specialized publications are likely to stem rather from foreign collaborations, while diversified ones from domestic extramuros collaborations. Results are not surprising since it is well known that especially in the life sciences, research projects frequently require a large number of research groups from all over the world, and significant amounts of resources that one single country might find it difficult to employ. In Physics these requirements are so intrinsic to the field that both international and domestic collaborations underlie specialized publications.

As for the robustness of outcomes, the proposed sensitivity analysis seems to indicate a truly satisfactory alignment of results varying thresholds for the dataset selection. Nevertheless, in any interpretation and for comparison in replicability of the analysis, we urge the reader to take account of the sensitivity of the results: i) to the convention adopted for definition of specialization/diversification in research activity; and in consequence ii) the field-classification scheme of the publications; iii) the field-classification scheme for the professors, as regards analyses at disciplinary level; iv) the observation period; and last but not least, v) the specific characteristics of the country system under analysis.

Future research could delve into the relatedness of fields in which the scientists diversify. Defining related those SCs classified within the same higher-levels categories (for example WoS or OECD categories),[11] and unrelated those which are not, it would be possible to investigate the relations between amplitude and type of collaboration (intramural, extramural domestic or international) and output of field-related and field-unrelated research diversification.

---

[11] See http://help.prod-incites.com/inCites2Live/filterValuesGroup/researchAreaSchema/oecdCategoryScheme.html, last access 20 February 2019

**APPENDIX - Robustness checks of the logistic regression**

**Scenarios definition**

Each dataset consists of professors whose 2012-2016 WoS-indexed publications meet the following requirements: i) at least 5 publications are WoS indexed; ii) publications fall in at least 2 different SCs; iii) publications present one prevalent SC; and iv):
- Scenario A: at least 50% of total publications (405,172 obs) fall in the prevalent SC.
- Scenario B: at least 70% of total publications (215,261 obs) fall in the prevalent SC.
- Scenario C: in the prevalent SC fall at least double as many publications as in the second SC (242,166 obs)
- Scenario D: in the prevalent SC fall at least triple as many publications as in the second SC (109,291 obs).

**Scenario A**

| UDA† | Obs | No. of authors | | cefp | | cedp | | cip | | Prob (χ2) |
|---|---|---|---|---|---|---|---|---|---|---|
| 1 | 19,310 | 1.001 | *** | 0.928 | ** | 0.751 | *** | 0.621 | *** | 0.000 |
| 2 | 89,078 | 1.000 | *** | 1.756 | *** | 1.093 | *** | 0.838 | *** | 0.000 |
| 3 | 30,716 | 1.000 | | 1.040 | | 0.939 | *** | 1.031 | | 0.027 |
| 4 | 8,262 | 0.997 | | 1.145 | *** | 1.012 | | 0.870 | ** | 0.000 |
| 5 | 28,838 | 1.001 | *** | 1.039 | | 0.906 | *** | 0.995 | | 0.000 |
| 6 | 116,260 | 1.000 | | 1.089 | *** | 0.924 | *** | 0.820 | *** | 0.000 |
| 7 | 23,082 | 1.000 | | 0.949 | * | 0.783 | *** | 1.096 | ** | 0.000 |
| 8 | 11,917 | 1.002 | *** | 0.952 | | 0.916 | ** | 0.754 | *** | 0.000 |
| 9 | 77,709 | 1.000 | *** | 1.018 | | 0.785 | *** | 0.919 | *** | 0.000 |

† *1 - Mathematics and computer science; 2 - Physics; 3 - Chemistry; 4 - Earth sciences; 5 - Biology; 6 - Medicine; 7 - Agricultural and veterinary sciences; 8 - Civil engineering; 9 - Industrial and information engineering.*
*Y=1 for specialized publication; Y=0 for diversified publication*
*\* $p < 0.1$; \*\* $p < 0.05$; \*\*\* $p < 0.01$*
*cp - publication in collaboration; cip - in intramural collaboration; cedp - in extramural domestic collaboration; cefp - in international collaboration.*

**Scenario B**

| UDA† | Obs | No. of authors | | cefp | | cedp | | cip | | Prob (χ2) |
|---|---|---|---|---|---|---|---|---|---|---|
| 1 | 11,489 | 1.000 | *** | 0.880 | ** | 0.720 | *** | 0.623 | *** | 0.000 |
| 2 | 69,484 | 1.000 | *** | 1.824 | *** | 0.926 | *** | 0.622 | *** | 0.000 |
| 3 | 9,542 | 0.978 | | 1.020 | | 1.058 | *** | 1.248 | | 0.001 |
| 4 | 2,561 | 0.996 | | 0.996 | *** | 0.957 | | 0.871 | ** | 0.407 |
| 5 | 8,406 | 1.000 | *** | 1.009 | | 0.930 | *** | 0.919 | | 0.445 |
| 6 | 52,637 | 1.000 | | 1.140 | *** | 0.895 | *** | 0.869 | *** | 0.000 |
| 7 | 10,895 | 0.996 | | 0.909 | * | 0.753 | *** | 1.140 | ** | 0.000 |
| 8 | 6,136 | 1.001 | *** | 0.912 | | 1.027 | ** | 0.776 | *** | 0.014 |
| 9 | 44,111 | 0.998 | *** | 0.997 | | 0.781 | *** | 0.923 | *** | 0.000 |

**Scenario C**

| UDA† | Obs | No. of authors | cefp | cedp | cip | Prob (χ2) |
|---|---|---|---|---|---|---|
| 1 | 8,975 | 1.001 *** | 0.918 | 0.649 *** | 0.479 *** | 0.000 |
| 2 | 43,658 | 1.000 *** | 2.050 *** | 1.072 | 0.742 *** | 0.000 |
| 3 | 19,089 | 1.001 * | 1.060 * | 0.929 ** | 1.110 ** | 0.002 |
| 4 | 3,762 | 1.018 *** | 1.102 | 0.855 ** | 0.825 ** | 0.000 |
| 5 | 24,308 | 1.001 *** | 1.121 *** | 0.869 *** | 0.890 *** | 0.000 |
| 6 | 92,858 | 1.000 | 1.151 *** | 0.926 *** | 0.814 *** | 0.000 |
| 7 | 15,654 | 1.001 | 0.958 | 0.744 *** | 1.082 | 0.000 |
| 8 | 3,806 | 1.002 ** | 0.900 | 0.810 *** | 0.830 ** | 0.001 |
| 9 | 30,056 | 0.999 *** | 0.973 | 0.650 *** | 0.896 ** | 0.000 |

**Scenario D**

| UDA† | Obs | No. of authors | cefp | cedp | cip | Prob (χ2) |
|---|---|---|---|---|---|---|
| 1 | 3,541 | 1.001 *** | 0.890 | 0.576 *** | 0.401 *** | 0.000 |
| 2 | 10,253 | 1.001 ** | 2.495 *** | 1.518 *** | 0.862 * | 0.000 |
| 3 | 7,075 | 1.000 | 0.995 | 0.929 | 1.042 | 0.651 |
| 4 | 1,378 | 1.023 ** | 1.255 * | 0.676 *** | 0.803 | 0.000 |
| 5 | 10,477 | 0.999 | 1.089 * | 0.843 *** | 0.909 | 0.000 |
| 6 | 56,299 | 0.999 *** | 1.160 *** | 0.898 *** | 0.809 *** | 0.000 |
| 7 | 7,530 | 1.001 * | 0.805 *** | 0.688 *** | 1.113 | 0.000 |
| 8 | 733 | 0.964 | 1.030 | 1.128 | 0.680 | 0.286 |
| 9 | 12,005 | 0.997 *** | 0.993 | 0.694 *** | 0.901 | 0.000 |